\documentclass[letter]{aa}
\usepackage{graphicx}
\usepackage{natbib}
\newcommand{\mdens}{{\rm g~cm^{-3}}}

\newcommand{\msun}{{\rm M}_\odot}
\newcommand{\mmax}{M_{\rm max}}
\begin{document}
\title{Two branches of neutron stars -
reconciling a $2{\rm M}_\odot$ pulsar
and SN1987A}

\author{P. Haensel\inst{1} \and M. Bejger\inst{1,2}
\and J.L. Zdunik\inst{1}
}
\institute{
N. Copernicus Astronomical Center, Polish
           Academy of Sciences, Bartycka 18, PL-00-716 Warszawa,
           Poland
           {\em jlz@camk.edu.pl}}
\institute{
N. Copernicus Astronomical Center, Polish
           Academy of Sciences, Bartycka 18, PL-00-716 Warszawa,
           Poland
           \and
           LUTh, Observatoire de Paris, CNRS, Universit{\'e} Paris
           Diderot, 5 Pl. Jules Janssen, 92190
            Meudon, France\\
           {\tt haensel@camk.edu.pl, bejger@camk.edu.pl,
           jlz@camk.edu.pl}}
%
\offprints{P. Haensel}
\date{Received xxx Accepted xxx}
\abstract{}{The analysis of SN1987A led Brown and  Bethe
(1995) to conclusion, that the maximum mass of cold
neutron stars is  low, $M_{\rm max}\approx 1.5\;{\rm M}_\odot$.
Such a  low  $M_{\rm max}$, due to a kaon condensation in the
stellar core, implies  collapse of a too massive deleptonized
protoneutron star into a black hole. This  would naturally
explain the lack of a neutron
star in the SN1987A remnant. On the other hand, recent
evaluation of mass of PSR J0751+1807 gives $M_{\rm max}\ga
2\;{\rm M}_\odot$. This contradicts the original Bethe-Brown
model, but can be reconciled within scenarios proposed in the
present Letter.}{We  consider two types of dense matter models
with high-density softening,  due to a transition from
a non-strange N-phase of matter to a
strangeness carrying phase S: kaon condensation and
deconfinement of quarks. Two scenarios of
neutron star formation in stellar core collapse are  considered. In the first scenario,
realized in sufficiently hot and dense supernova cores,
nucleation of an S-phase is sufficiently rapid so as to
form an  S-phase core, and implying
$M_{\rm max}=M_{\rm max}^{\rm S}\approx 1.5\;{\rm M}_\odot$.
 In the second scenario, nucleation of the S-phase at neutron star birth is
too slow to materialize, and the star becomes cold without
forming an S-phase core. Then, stellar mass can increase
via accretion, until central density $\rho_{\rm crit}$ is
reached, and the S phase forms. This N branch of neutron stars ends at
$M=M_{\rm crit}$.}{ We select  several
models of N-phase satifying the necessary
condition $M^{\rm N}_{\rm max}\ga 2\;{\rm M}_\odot$ and
combine them with models of kaon condensation and quark
deconfinement. For kaon condensation, we get $M_{\rm
crit}\approx M^{\rm S}_{\rm max}\approx 1.5\;{\rm M}_\odot$,
 which is ruled out by PSR J0751+1807.
On the contrary, for the EOSs with
 quark deconfinement we get
$M_{\rm crit}\approx M^{\rm N}_{\rm max}\ga 2\;{\rm
M}_\odot$, which reconciles SN1987A and PSR J0751+1807.}{}

\titlerunning{Two branches of neutron stars}
\maketitle
%
\section{Introduction}
\label{sect:introd}
The actual equation of state (EOS) is one of the secrets of
neutron stars. Uncertainties of our models of neutron star
cores at density exceeding significantly the normal nuclear
density $\rho_0=2.7\times 10^{14}~\mdens$ results in the
ignorance about the EOS at, say, $\rho\ga 2\rho_0$. An
important characteristics of an EOS model  is the maximum
allowable mass of neutron stars that it predicts, $M_{\rm
max}$. Mathematically, $M_{\rm max}$ is a functional of the
EOS, $M_{\rm max}=M_{\rm max}[P(\rho)]$. Rotation increases
$M_{\rm max}$, but even for highest observed pulsar frequency
this increase is $\sim 3\%$ and can therefore be neglected
(see, e.g., \citealt{NSB1}).

Measured neutron star masses increase in number and in
precision (for a recent review, see \citealt{NSB1}). Let us
denote the largest measured neutron star mass by $M_{\rm
obs}^{\rm max}$. To be consistent with observations, an EOS
should yield $\mmax> M_{\rm obs}^{\rm max}$. Therefore,
higher the value of $M_{\rm obs}^{\rm max}$, stronger the constraint on the
stiffness of the EOS of neutron star core. Masses
significantly higher than $1.4\;{\rm M}_\odot$ were measured
in some pulsar - white dwarf binaries, the largest of them was
$2.1\pm 0.2\;{\rm M}_\odot$ for PSR J0751+1807
(\citealt{Nice2005}). Clearly, $M_{\rm max}\ga 2\;{\rm M}_\odot$
necessitates a  stiff EOS.

This conclusion seems to be in conflict with a puzzling
absence of a neutron star in the SN1987A remnant. Indeed,
\cite{BB1995} concluded, that $0.075{\rm M}_\odot$ of the
radioactive $^{56}{\rm Ni}$ produced by SN1987A together with
other characteristics of this SN and of the presupernova star,
imply an upper bound $M_{\rm max}\approx   1.5\;{\rm M}_\odot$.
Such a low value of $M_{\rm max}$ has been attributed by
\cite{BB1995} to the kaon condensation in neutron star core,
and was also used as an argument for a large number
of  low-mass black holes in the Galaxy (\citealt{BB1994}).

In the present Letter we show that  a low
upper bound $\approx 1.5\;\msun$ from SN 1987A, and a high lower bound $\ga
2\;\msun$ from the mass of PSR J0751+1807, can be reconciled
 if they refer to two different branches of neutron stars,
   formed in different evolutionary  scenarios. The
low-$\mmax$ branch consists of configurations with
superdense strangeness-carrying (S) cores and was
assumed to be formed in a collapse
of massive stars ($20-30\;\msun$ on the main sequence).
The S-phase, characterized by a
significant strangeness per baryon, is assumed to nucleate
at  sufficiently high temperature and density  in
a newborn neutron star. The S phase core remains in equilibrium with
a non-strange (or a weakly strange) N-phase envelope.
 Such configurations form an S-branch of neutron stars, with $M_{\rm
max}^{\rm S}\approx 1.5\;\msun$. On the other hand, the
configurations laying on the high-$\mmax$ branch
are assumed to be initially  formed
as low-mass N-phase stars, and reached their
present state by accretion of matter in
a long-living close binary system. Such stars can nucleate the
S-phase only after reaching a critical mass
$M_{\rm crit}\ga 2\;\msun$: massive neutron stars with white dwarf
companions were formed in such a way. The notion of
$M_{\rm crit}$ as a maximum mass for a branch
of neutron stars, metastable with respect to formation of
a quark core,  was recently studied in detail by
\cite{Bombaci2007} (see also \citealt{Bombaci2004}). We give specific
examples of dense matter models with N-S phase transition, and
we show, that the original proposal of \cite{BB1995} where the
S-phase resulted from kaon condensation fails to produce a
sufficiently high $M_{\rm crit}$. However, the \cite{BB1995}
bound from SN1987A can be reconciled with ~$2\;\msun$
neutron stars, provided they
could follow  different formation and evolution tracks,
proposed in the present Letter.

In Sect.\ \ref{sect:EOSs} we briefly describe the EOSs
involving kaon condensation and quark decofinement. Two
branches of neutron stars, for kaon-condensed and
quark-matter cores, are constructed in  Sect.\
\ref{sect:Branches}. Astrophysical scenarios leading to two neutron
star branches  are summarized in Sect.\ \ref{sect:twoMmaxs}.
\section{EOSs with phase transitions: stable and metastable branches}
\label{sect:EOSs}
\subsection{Kaon condensation}
\label{sect:EOS_NK}
Interaction with nucleons can greatly reduce the energy of a
single ${K}^-$ in nuclear matter, $\omega_K$. Consequently,
kaons could appear in nucleon matter, forming a Bose-Einstein
condensate (\citealt{KaplanNelson1986}; for review see
\citealt{Ramos2001}). In what follows, we consider kaon
condensation in neutron star cores composed of neutrons,
protons, electrons, and muons ($npe\mu$ matter). Nucleons are described using
two versions of the  relativistic mean-field model with scalar
self-coupling (\citealt{GM1991}, \citealt{ZM1990}). Coupling of kaons to nucleons is
described by the models of \cite{GlendSB1999}, the crucial
parameter being the depth of the potential well of kaons in
symmetric nuclear matter at saturation point, $U^{\rm lin}_K$.
In what follows we will measure temperature $T$ in MeV
($T[{\rm MeV}]=0.8617\cdot T[{\rm K}]/10^{10}$). We
will consider two basic cases. The low $T$ case corresponds
to $T<1\;$MeV. Such conditions prevail in the cores
of single neutron stars older than one day, as well as
in the neutron stars accreting matter in binary systems.
The high $T$ case with $T\approx 50\;$MeV can be realized
in the central core of newborn neutron star during
10 s just after the deleptonization.
\begin{table}
\caption{Densities $\rho{\rm _N}$ and $\rho_{\rm crit}$
 and maximum neutron star masses. {\it Upper panel: EOSs with
kaon condensation.} The first two letters refer to the nucleon EOS
(ZM - \citealt{ZM1990}; GM - \citealt{GM1991}) and the last two
denote the kaon condensation model (GS - \citealt{GlendSB1999}).
The three digit number gives the value of
$-U^{\rm lin}_K$. {\it Lower panel: EOSs with quark
deconfinement.}
Letters denote the nucleon EOS: APR - \cite{APR1998}; GN -
case 5 in \cite{Glend1985}. Quark matter is described
within the MIT Bag Model (see, e.g., \citealt{FJ1984}).  First number  gives the
the value of the bag constant (in ${\rm MeV~ fm^{-3}}$), while the
second number gives the mass of the
$s$ quark (in MeV). In all cases the QCD coupling
constant $\alpha_{\rm s}=0.2$.  For further explanations see the text.}
\begin{center}
\begin{tabular}[t]{|ccccc|}
\hline
 model & ${\rho_{\rm _N}}^a$ &
 ${\rho_{\rm  crit}}^a$ &
 ${M^{\rm S}_{\rm max}}^b$  & ${M_{\rm crit}}^b$ \\
\hline
 ZMGS100 & 8.78 & 9.14 & 1.57 & 1.55 \\
 GMGS130 & 8.13 &   8.85 & 1.42 & 1.53\\
 \hline
 \hline
  APR.100.150 & 10.3 & 20.1 & 1.57& 2.13 \\
 APR.90.200 &  11.0 &   19.7 & 1.62 & 2.13\\
  GN.100.150 &  5.06 &   22.0 & 1.48 & 2.145$^c$\\
   GN.90.200 &   5.14 &   20.9 & 1.51 & 2.14\\
   \hline
 \end{tabular}
 \end{center}
   $^a$ in  $\rm 10^{14}~{\rm g~cm^{-3}}$, $^b$
  in $\msun$. $^c$ Here $\rho_{\rm crit}>\rho_{\rm c,max}$ and therefore
   $M_{\rm crit}=M^{\rm N}_{\rm max}$.
 \end{table}

\begin{figure}[h]
\centering \resizebox{3.3in}{!}{\includegraphics[angle=-90]{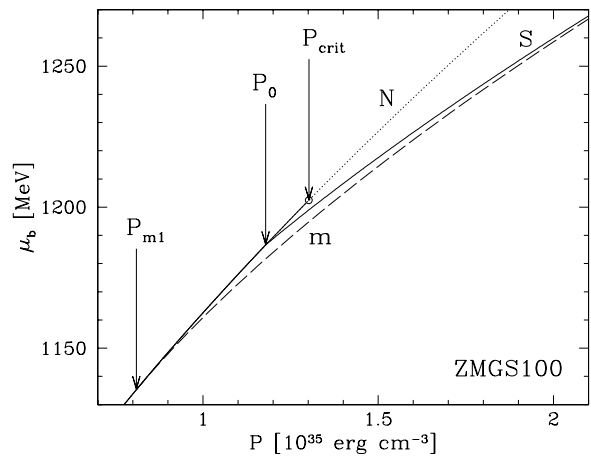}}
\caption{Baryon chemical potential $\mu_{\rm b}=({\cal
E}+P)/n_{\rm b}$ (${\cal E}$ - energy density, $n_{\rm b}$ -
baryon number density) vs. pressure. $T=0$ approximation
is assumed, calculations for the ZMGS100 model
of kaon condensation in $npe\mu$ matter (see Table 1).
 Dotted line - unstable N phase. Long dashes -
mixed phase.}
 \label{fig:muNK}
\end{figure}

Let us first assume $T<1\;$MeV.  All
core constituents are strongly degenerate. The $npe\mu$ matter
becomes {unstable} with respect to
spontaneous formation of kaons at density $\rho_{\rm crit}$,
at which  the electron Fermi energy $\mu_e=\omega_K$. Then the
density of kaons grows till the equilibrium between the normal
N phase of dense matter and the kaon-condensed S phase
is reached. Two possibilities of the coexistence of the N and S phases
should be contemplated. The coexistence of the pure N and S phases takes
place at a well defined pressure, $P_0$,  and is associated with a
density jump $\rho_{\rm _N}\longrightarrow \rho_{\rm  _S}$, or it occurs
via a mixed-phase state in the pressure interval $P_{\rm m1}<P<P_{\rm
m2}$. Whether the phase transition takes place at a constant pressure,
or via a mixed state, depends on the poorly known surface
tension at the N-S interface, and on the scenario of formation
of a kaon condensed  core. An example of baryon chemical
potential vs. pressure for different phases is shown in Fig.\
\ref{fig:muNK}, and the parameters associated with the
${\rm N}\longrightarrow {\rm S}$ transition models are given in Table 1.
In the high-$T$ case the
inclusion of the $T$ dependence of the EOS is mandatory. Moreover,
high $T$ makes the kaon condensation weaker (\citealt{PonsKcond2000}).
However, the  thermal effects do not affect our conclusions regarding
cooled neutron stars.
\vskip -0.5cm

\begin{figure}[h]
\centering \resizebox{3.3in}{!}{\includegraphics[angle=-90]{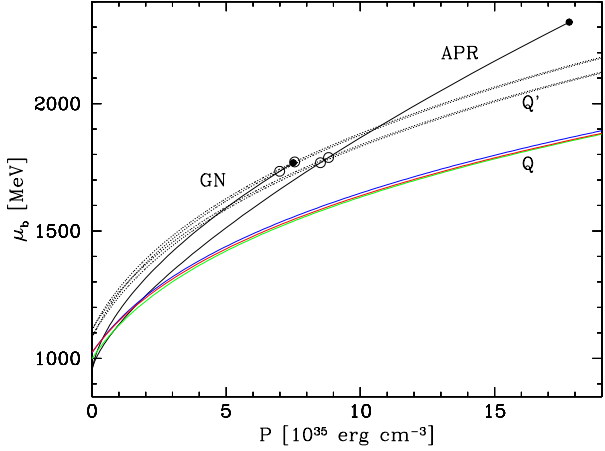}}
\caption{(Color online). Baryon chemical potential vs.
pressure. Solid lines - two models of the $npe\mu$ matter, filled circles -
maximum central density of neutron stars
built of the N phase. Open circles  -
first-order phase transition to non-strange quark matter.
Parameters  $B$, $\alpha_{\rm s}$, and $m_s$  of quark matter are
the same as in Table 1.
}
 \label{fig:mu_NQ}
\end{figure}

\begin{figure}[h]
\centering \resizebox{3.4in}{!}{\includegraphics[angle=-90]{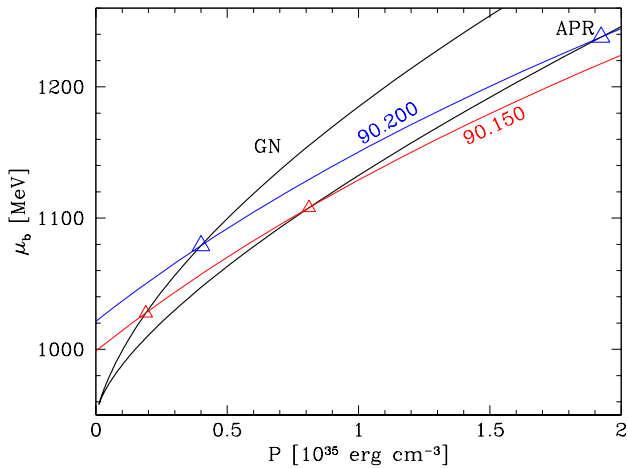}}
\caption{(Color online). Phase transitions N$\longrightarrow$Q to
three-flavor quark matter in beta and strangeness equilibrium, at $T=0$.
APR and GN models of the N phase are used.
Models of Q phase are labeled by the values of $B$ and $m_{\rm s}$.
In both cases $\alpha_{\rm s}=0.2$. Open triangles - first-order phase
transition to quark matter.
}
 \label{fig:mu_NQ_zoom}
\end{figure}
\vskip -0.5cm
\subsection{Deconfinement of quarks}
\label{sect:EOS_NQ}
Here the N phase is again the $npe\mu$ matter. Quark deconfinement
is a strong interaction process. Therefore, it
produces the non-strange quark plasma of
the $u$ and $d$ quarks,  with the same lepton fractions
(per baryon) as in the N phase. This quark plasma state will be
called ${\rm Q}^\prime$. The
transition ${\rm N}\longrightarrow {\rm Q}^\prime$ takes place at some
$P_{\rm crit}$. Then, weak interactions convert
nearly half of the $d$ quarks into the $s$ ones, producing the
equilibrated $uds$ quark plasma state ${\rm Q}$.  The quark phases
${\rm Q}^\prime$ and Q are described using the MIT Bag model,
with parameters given in Table 1. In all cases, we use the
same QCD coupling constant $\alpha_{\rm s}=0.2$.

In Fig.\ \ref{fig:mu_NQ} we plotted baryon chemical potential
$\mu_{\rm b}=({\cal E}+P)/n_{\rm b}$ versus pressure $P$, for two models
of the N phase, and several models of the ${\rm Q}^\prime$ and Q phases
of quark matter. At given $P$, the ${\rm Q}^\prime$ phase has the same
lepton fractions, $x_e=n_e/n_{\rm b}$ and $x_\mu=n_\mu/n_{\rm b}$, as the beta
equilibrated N phase.  As far as the quark matter phases are concerned, some
generic features can be pointed out. For a given model of the N phase, and a fixed
value of $\alpha_{\rm s}$, higher value of $B$ gives a higher value of
$P_{\rm crit}$. On the other hand, for a given pair $B$, $\alpha_{\rm s}$,
the pressure $P_{\rm crit}$ for the softer EOS (APR) is higher than for the stiffer
one (GN). After strangeness and beta equilibration, the phase transition
pressure, $P_0$, is much lower than $P_{\rm crit}$.
The dependence of $P_0$ on the quark matter model parameters is quite
strong, as seen in Fig.\ \ref{fig:mu_NQ_zoom}.

For $T<1~$MeV  all dense matter
phases are strongly degenerate, and phase transition
parameters  can be very well approximated by the values
obtained at $T=0$. An example of pressure dependence of the
chemical potentials, and resulting phase transition points, is
given in Fig.\ \ref{fig:mu_NQ}.  Nucleation of the ${\rm
Q}^\prime$ phase is realized there in the quantum regime via
energy barrier penetration, at the pressure slightly larger
than $P_{\rm crit}$  (see \citealt{Bombaci2007} and references
therein).

For $T\approx 50\;$MeV, transition to the
 ${\rm Q}^\prime$ phase occurs at significantly lower density
than $\rho_{\rm crit}(T=0)$ (see, e.g., \citealt{Lugones1998}).
 Moreover, at $T\ga 50 \;$MeV thermal
nucleation of the ${\rm Q}^\prime$ phase,
via jumping over the energy barrier separating
the N and ${\rm Q}^\prime$ states,  is very efficient.
Additionally, at such a high $T$ a Boltzman gas of lightest
hyperons is present (even if it is absent at $T<1\;$MeV).
These thermal hyperons act as nucleation centers for the quark
matter, for they add  the $s$ quarks to the
${\rm Q}^\prime$ phase. A high $T$ greatly accelerates
$u+d\longrightarrow s+u$ process, leading to a rapid
  strangeness equilibration
associated with ${\rm Q}^\prime\longrightarrow {\rm Q}$.
\vskip -0.5cm
\section{Two branches of neutron stars }
\label{sect:Branches}
\vskip -0.7cm
\begin{figure}[h]
\centering \resizebox{3.4in}{!}{\includegraphics[angle=-90]{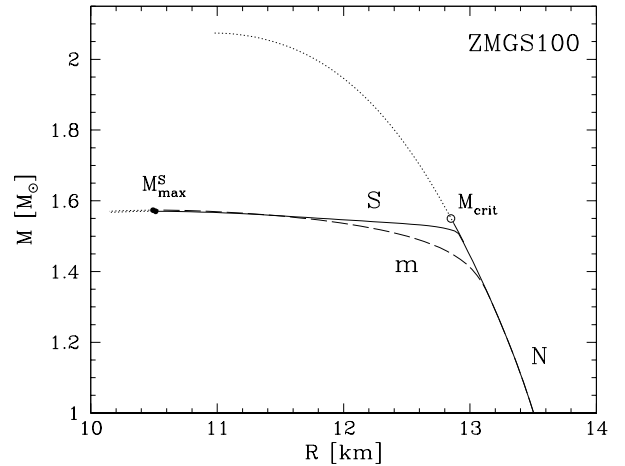}}
\caption{Branches $M(R)$ for EOSs with kaon condensation.
Configurations on the dotted
lines are unstable. For
further explanation see the text.
}
 \label{fig:MR_NK}
\end{figure}
\vskip -0.5cm
Let us start with kaon condensation.
The $M(R)$ plots at $T<1\;$MeV, obtained by solving  the
Tolman-Oppenheimer-Volkov equations of hydrostatic equilibrium
(see, e.g., \citealt{NSB1}),  are shown in Fig.\
\ref{fig:MR_NK}. The N-branch of cold stars,
built exclusively of N phase, ends at $M_{\rm crit}=M^{\rm N}(\rho_{\rm c}=\rho_{\rm
crit})$. There are two branches with S phase cores. The S
branch consists of configurations with a
 pure S phase core, while configurations on the
m branch have  mixed-phase cores. However, as one can see, they both end at
(nearly) the same maximum mass, denoted by $M^{\rm S}_{\rm max}$.

There are two basic formation scenarios. If a newborn hot neutron
star did not nucleate kaon condensate, then after cooling it
lies on the N branch. By accreting matter in a close binary
system, such a neutron star can move upwards along the N
branch, up to $M_{\rm crit}$, and then collapses into a black
hole (case GMGS130 in Table 1) or develops a kaon condensed
core and settles on the S branch or m branch (case ZMGS100 in
Table 1). However, in all cases  the maximum mass $M_{\rm
crit}\approx M^{\rm S}_{\rm max}\approx 1.5\;\msun$.
\vskip -0.5cm
\begin{figure}[h]
\centering \resizebox{3.3in}{!}{\includegraphics[angle=-90]{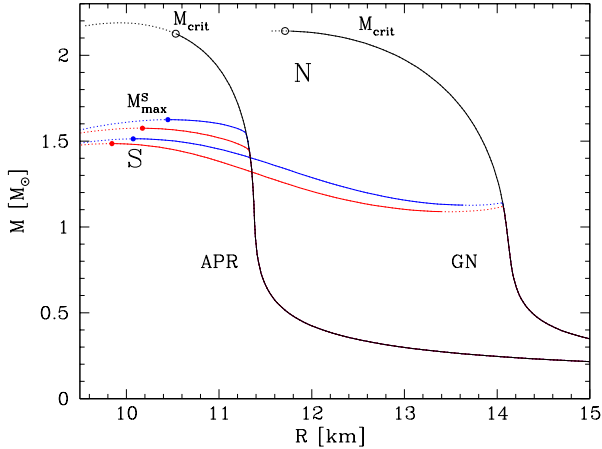}}
\caption{(Color online) Branches $M(R)$ for EOSs with quark deconfinement
phase transition (Table 1). Configurations on the dotted segments
are unstable. Open circle: configuration of N phase with
$M=M_{\rm crit}$. Filled circle: limiting configuration with S phase core
and  $M=M^{\rm S}_{\rm max}$.
 For further explanation see the text.
}
 \label{fig:MR_NQ}
\end{figure}
Consider now the case of quark deconfinement.
The $M(R)$ plots for several models of the N, ${\rm Q}^\prime$, and Q
phases  are displayed in Fig.\ \ref{fig:MR_NQ}. Two branches
of neutron stars are clearly visible. The S branch shape, and
in  particular its initial point at central density
$\rho_{\rm c}=\rho_{\rm _N}$, are quite sensitive to the dense
matter model. However, the maximum mass of the S branch, i.e.
$M_{\rm max}^{\rm S}$, is quite independent of the dense matter
model, $M_{\rm max}^{\rm S}\approx 1.5\;\msun$. In our calculations,
the quark cores of configurations on the S branch are made of pure quark
matter, but we know from \cite{Alford2006},  that the value of
$M_{\rm max}^{\rm S}$, calculated for the mixed quark-nucleon
cores, is to a very good approximation the same as for the pure
quark-matter cores. The  N branches end at
$\rho_{\rm c}=\rho_{\rm crit}$, at $M_{\rm crit}\ga
2\;\msun$. Let us stress that the N branch configurations with
$M>M^{\rm S}_{\rm max}$ {\it have no twins} of the same
mass on the  S branch. Rather, they are ``high mass siblings''
of ``low mass'' neutron stars of  branch S.

Consider a low $T$ scenario.
A newborn neutron star had, after deleptonization,
central $T$ and $\rho$ too low to  nucleate the ${\rm
Q}^\prime$ phase during some 10 s of extremal central
conditions. After cooling, the star will lie on the N branch,
Fig.\ \ref{fig:MR_NQ}, and can move upwards along this
branch due to accretion.

The high-$T$ scenario is different. We assume that
during critical 10 s after deleptonization,
central temperature $T\ga 50\;$MeV, and  central density
exceeds $\rho_{\rm crit}(T)$. Consequently, ${\rm Q}^\prime$
phase forms and transforms into the Q phase by strangeness-changing
reaction. Two final states are then possible. Either neutron star
settles  on the S branch, Fig.\ \ref{fig:MR_NQ},
with maximum mass $M^{\rm S}_{\rm max}\approx
1.5\;\msun$, or it collapses into a black hole if its mass
$M>M^{\rm S}_{\rm max}$. If  the star settles on the S
branch, further mass accretion can move it upwards along this
branch. After reaching  $M^{\rm S}_{\rm max}$, the star
will collapse into a black hole.
\vskip -1.cm
\section{Two  scenarios resulting in two maximum  masses}
\label{sect:twoMmaxs}
The original Brown-Bethe (BB) scenario involved kaon
condensation in a deleptonized newborn neutron star,
and predicted  maximum mass of $1.5\;\msun$ for cold neutron stars.
The BB scenario applied to type
II SNae, originating  from  collapse of massive stars (main-sequence mass
$20-30\;\msun$). This scenario  could explain
the absence of neutron star in the SN1987A remnant and
the production of $0.075\;\msun$ of radioactive $^{56}{\rm Ni}$ in
that supernova. We have shown that even neutron stars that
initially  had no  kaon condensed core, and increased their mass by accretion,
  could never exceed  $M_{\rm crit}\approx 1.5\;\msun$; this would contradict
  recent measurements of  neutron stars masses. We argue that the BB
scenario could be saved, if kaon condensation, causing the
softening of the EOS,  is replaced by the
quark deconfinement.  Quark deconfinement might allow
for  two scenarios associated with two significantly
 different maximum masses.
In the first one, neutron stars, born in core-collapse of
massive stars ($20-30\;\msun$), were sufficiently hot and
dense after deleptonization to produce EOS-softening quark
core, which resulted in $M^{\rm S}_{\rm max}\approx
1.5\;\msun$; this was the case of SN1987A. However, there are
also  type II SNae, resulting from the collapse of less massive
stars, when a hot deleptonized neutron star has no quark core.
This could be the case of SNae produced by the electron-capture
collapse of degenerate O-Ne-Mg cores of helium stars
(\citealt{vdHeuvel2007} and references therein). After cooling,
such neutron star could increase its mass by accretion in a
long-lived binary system, up to $M_{\rm max}\ga
2\;\msun$. This second maximum mass, characteristic of the
normal (no quark core) neutron-star branch, is just a
critical mass for nucleation of a low-strangeness quark matter.
It is consistent with largest measured masses of pulsars with white
dwarf companions.
\vskip -1.cm
\acknowledgements{ This work was partially
supported by the Polish MNiI grant no. 1P03D.008.27. MB was
also partially supported by the Marie Curie Intra-european
Fellowship MEIF-CT-2005-023644 within the 6th European
Community Framework Programme.}


\vskip -1.cm
\begin{thebibliography}{}
{\scriptsize
%
\bibitem[Alford et al.(2006)]{Alford2006}
Alford M., Braby M., Paris M., Reddy S., 2006,
ApJ, 629, 969
%
\bibitem[Akmal et al.(1998)]{APR1998}
Akmal A., Pandharipande V.R., Ravenhall D.G., 1998, Phys.Rev.
C, 58, 1804
%
\bibitem[Bethe \&  Brown (1995)]{BB1995}
Bethe H.A., Brown G.E., 1995, ApJ, 445, L129
%
\bibitem[Brown \& Bethe (1994)]{BB1994}
Brown G.E., Bethe H.A., 1994, ApJ, 423, 659
%
%
\bibitem[Bombaci et al.(2004)]{Bombaci2004}
Bombaci I., Parenti I., Vida${\widetilde{\rm n}}$a, 2004,
A \& A, 462, 1017
%
\bibitem[Bombaci et al.(2007)]{Bombaci2007}
Bombaci I., Lugones G., Vida${\widetilde{\rm n}}$a, 2007,
A \& A, 462, 1017
%
\bibitem[Farhi \& Jaffe (1984)]{FJ1984}
Farhi E., Jaffe R.L., 1984,
Phys. Rev. D, 30, 2379
%
\bibitem[Glendenning(1985)]{Glend1985}
Glendenning, N. K. 1985, ApJ, 293, 470
%
\bibitem[Glendenning \& Moszkowski (1991)]{GM1991}
Glendenning, N. K., Moszkowski S.A., 1991, Phys. Rev. Lett.,
67, 2414
%
\bibitem[Glendenning \& Schaffner-Bielich (1999)]{GlendSB1999}
Glendenning N.K., Schaffner-Bielich J.,
1999, Phys. Rev. C, 60, 025803
%
\bibitem[Haensel et al. (2007)]{NSB1}
Haensel P., Potekhin A.Y., Yakovlev D.G., 2007
Neutron Stars 1. Equation of State and Structure,
(Springer, New York)
%
\bibitem[van den Heuvel (2007)]{vdHeuvel2007}
van den Heuvel E.P.J, 2007, {\tt arXiv:0704.1215[astro-ph]}
%
%
%
%
\bibitem[Kaplan \& Nelson (1986)]{KaplanNelson1986}
Kaplan  J.L., Nelson A.E., 1986, Phys. Lett. B, 175, 57
%
\bibitem[Lugones \& Benvenuto (1998)]{Lugones1998}
Lugones G., Benvenuto O.G., 1998,
Phys. Rev. D, 58, 083001
%
\bibitem[Nice et al.(2005)]{Nice2005}
Nice D., Splaver E.M., Stairs I.H., et al., 2005,
 ApJ, 634, 1242
%
\bibitem[Pons et al.(2000)]{PonsKcond2000}
Pons J.A., Reddy S., Ellis P.J., Prakash M., Lattimer J.M.,
2000, Phys. Rev. C, 62, 035803
%
\bibitem[Ramos  et al.(2001)]{Ramos2001}
Ramos A., Schaffner-Bielich J., Wambach J., 2001, in: Physics of Neutron Star
Interiors, ed. by D. Blaschke, N.K. Glendenning, \& A. Sedrakian,
Lecture Notes in Phys., 578, 175
%
\bibitem[Zimanyi \& Moszkowski (1990)]{ZM1990}
Zimanyi J., Moszkowski S.A., Phys. Rev. C, 42,1416
%
}
\end{thebibliography}
\end{document}